# Superconducting LaAlO$_3$/SrTiO$_3$ Nanowires

Joshua P. Veazey[1], Guanglei Cheng[1], Patrick Irvin[1], Cheng Cen[1,†], Daniela F. Bogorin[1,‡], Feng Bi[1], Mengchen Huang[1], Chung-Wung Bark[2], Sangwoo Ryu[2], Kwang-Hwan Cho[2], Chang-Beom Eom[2] and Jeremy Levy[1,*]

[1]Department of Physics and Astronomy, University of Pittsburgh, Pittsburgh, Pennsylvania 15260, USA

[2]Department of Materials Science and Engineering, University of Wisconsin-Madison, Madison, Wisconsin 53706, USA

**Abstract:** We report superconductivity in quasi-1D nanostructures created at the LaAlO$_3$/SrTiO$_3$ interface. Nanostructures having line widths $w$~10 nm are formed from the parent two-dimensional electron liquid using conductive atomic force microscope lithography. Nanowire cross-sections are small compared to the superconducting coherence length in LaAlO$_3$/SrTiO$_3$ ($w \ll \xi_{SC}$~100 nm), placing them in the quasi-1D regime. Broad superconducting transitions with temperature and finite resistances in the superconducting state well below $T_c \approx 200$ mK are observed. *V-I* curves show switching between the superconducting and normal states that are characteristic of superconducting nanowires. The four-terminal resistance in the superconducting state shows an unusual dependence on the current path, varying by as much as an order of magnitude.

Superconductivity in the quasi-one-dimensional (q-1D) regime is of diverse and fundamental interest [1]. The implementation of superconducting nanoelectronics for quantum information processing has additional practical appeal [2, 3]. In recent years, Majorana zero-modes in topological superconductors have received a great deal of attention theoretically [3, 4], particularly in one dimension [3, 5]. Nonlocal entanglement of zero-modes is seen as a path toward fault-tolerant quantum computation. One class of proposals [5-7] for observing Majorana fermions in the solid-state requires several basic ingredients: a 1D semiconducting nanowire with strong spin-orbit coupling, magnetic Zeeman field, and an *s*-wave superconductor proximity-coupled to the nanowire. Implementations of this basic outline have led to recent landmark experimental successes [8, 9].

The two-dimensional electron liquid (2DEL) at the $LaAlO_3/SrTiO_3$ interface possesses all of the necessary individual ingredients [10-13] to host Majorana zero-modes [14]. $SrTiO_3$ is one of a few so-called "superconducting semiconductors" with properties shared by both classes of materials [15]. In addition, the superconductivity coexists with ferromagnetism [16-18]. It remains an open question whether the various interactions between spin-orbit coupling, superconductivity, magnetism and nanoscale confinement will be suitable for supporting Majorana bound states. Here, our study is focused on fundamental studies of quasi-one-dimensional superconductivity in oxide nanostructures.

To create superconducting $LaAlO_3/SrTiO_3$ nanowires, we employ conductive atomic force microscopy (c-AFM) lithography [19, 20] to produce structures with line widths $w$~10 nm. Below $T_c \approx 200$ mK, the superconducting 2DEL is confined to a thickness estimated to be between $t$=1-10 nm [10, 21]. Nanowires are therefore expected to have cross sections that are small compared to the superconducting coherence length ($w, t \ll \xi_{SC}$ ~ 100 nm) [10, 22], placing them in the quasi-1D regime. The ability to "write" superconducting nanostructures on an insulating LaAlO$_3$/SrTiO$_3$ "canvas" opens possibilities for the development of new families of superconducting nanoelectronics.

LaAlO$_3$/SrTiO$_3$ heterostructures, grown by pulsed laser deposition, consist of 3.4 unit cells of LaAlO$_3$ on TiO$_2$-terminated SrTiO$_3$. Samples from two similar growth protocols (temperature $T$ and partial oxygen pressure $P_{O_2}$) were used: $T$=550 °C and $P_{O_2} = 10^{-3}$ mbar, or $T$=780 °C and $P_{O_2} = 7.5 \times 10^{-5}$ mbar. The transport properties described here do not show any obvious dependence on growth parameters, and essential features are observed for both sample types. Electrical contact to the interface where the 2DEL is formed is established by etching the LaAlO$_3$ to expose the SrTiO$_3$ and subsequently depositing 4 nm Ti and 30 nm Au. The LaAlO$_3$/SrTiO$_3$ nanostructures are then written at the interface using the c-AFM lithographic technique described in detail elsewhere [19], and consist of five- or six-terminal Hall bars [Fig. 1(a)] with main channel lengths $L$ between 2 and 10 μm. Nanostructures are constructed from nanowire segments sketched with +10—15 V applied to the tip, resulting in widths of $w$~10 nm [19, 23]. Table I summarizes properties of several devices studied.

Magnetotransport measurements are carried out with the sample placed in a dilution refrigerator and magnetic field directed out-of-plane. Four-terminal voltage vs. current curves, $V^{cd}(I^{ab})$, are acquired by applying a voltage $V^a$ at electrode $a$ and measuring current $I^{ab}$ at electrode $b$, which is held at virtual ground. Simultaneously, the voltage difference $V^{cd} = V^c - V^d$ between electrodes $c$ and $d$ is measured with a ~TΩ input impedance differential amplifier. In all device diagrams marking current pathway $I^{ab}$, the arrows point from leads $a$ to $b$. The differential resistance $R^{ab,cd} \equiv dV^{cd}/dI^{ab}$ is obtained by numerical differentiation. We use $R_C^{ab,cd}$ to denote resistance below $I_c$, while $R_N^{ab,cd}$ denotes resistance above $I_c$. Resistances are expressed in units of the resistance quantum ($h/e^2$), where $h$ is the Planck constant and $e$ is the electron charge. We define the critical temperature $T_c$ as the temperature at which the resistance has dropped by half the difference between the normal ($T \gg T_c$) and superconducting ($T \ll T_c$) resistances. The carrier densities measured for these nanostructures ($n \approx 1.7 - 2.4 \times 10^{13} \text{cm}^{-2}$) from Hall magnetoresistance fall within a range where superconductivity and strong spin-orbit coupling coexist in 2D geometries [10, 11, 24].

LaAlO$_3$/SrTiO$_3$ nanowires routinely have resistances that remain finite in the superconducting state at temperatures well below the transition. The $V$-$I$ and d$V$/d$I$ characteristics of Device N2 are shown in Fig. 1(b) and 1(c). In the superconducting phase of Device N2, $V^{23}(I^{14})$ has a finite slope, and $R_C^{14,23} = 7.0 \times 10^{-2} h/e^2$ (1.8 kΩ) below the critical current $I_c = 7$ nA. When $I > I_c$ the normal resistance is $R_N^{14,23} = 4.0 \times 10^{-1} h/e^2$ (10.3 kΩ). The superconducting phase in Device N2 disappears for sufficiently elevated magnetic fields, above $|H| \approx 2$ kOe or temperatures, $T > T_c = 110$ mK. As the temperature is decreased through the superconducting transition [Fig. 2], the peaks in d$V$/d$I$ gradually widen, showing a broad transition width in temperature. The full temperature and magnetic field dependence in these devices indicate critical temperatures and magnetic fields that are consistent with those of previously reported planar structures [10, 11]. See Supplemental Material for a phase diagram with temperature and magnetic field for Device N3.

Figure 3 shows the $V$-$I$ characteristics for three devices: N2, N3, and N4 at T=50 mK. Rather than a sudden voltage increase at a critical current $I_c$, the switching currents in these nanostructures have a finite widths. Here, $I_c$ is defined as the location of the peak in d$V$/d$I$ between superconducting and normal states [see Fig. 1(c)]. Finite widths in switching current are sometimes observed in Josephson junctions (JJ). Superconducting nanowires may be modeled as JJ of the form *S-c-S*, where *S* is a superconducting reservoir and *c* is a constriction in the superconductor [25]. The constriction occurs at either a nanowire segment or the nanowire itself. While critical currents ($I_c \approx 3-9$ nA) and normal-state resistances in LaAlO$_3$/SrTiO$_3$ nanostructures vary by device, the products $I_c R_N$ agree well with the *S-c-S* characterization ($I_c R_N \sim k_B T_c / e$), where $k_B$ is the Boltzmann constant [26].

The *V-I* curves in Fig. 3 also reveal the functional dependence for the low-resistance state as the source current *I* is increased from zero toward $I_c$. While planar LaAlO$_3$/SrTiO$_3$ structures when *T*<<*T*$_c$ have linear *V-I* curves in the superconducting state until a sharp voltage jump at *I*=$I_c$ [10], the behavior in the nanostructures is quite different. The *V-I* dependence becomes non-linear for currents approaching *I*~$I_c$. As the current crosses through $I_c$, the four-terminal voltage rises much more rapidly but the transition has a finite width. In Device N4, the current reaches a maximum (defined as $I_c$) and then decreases again as the voltages continues to rise. This device also shows a hysteresis when the source current is swept in opposite directions [inset, Fig. 3]. Device N2 and N3 have five terminals and L=10 μm, while Device N4 has six terminals and L=6.5 μm. However, for all devices studied, we observe no systematic dependence in *V-I* or other properties on the number of leads or channel length [Table I].

The four-terminal resistance in the superconducting state exhibits a surprising dependence on the configuration of the voltage and current leads. For example, simply changing the current pathway changes $R_C$ by a factor of three in Device N2 [Fig. 4]. However, the voltage leads are fixed and the segment where the voltage drop is measured remains unchanged. In another configuration where current and voltage leads are swapped, $R_C$ drops by an order of magnitude [Fig. 4]. The normal state resistance $R_N$ is the same in all three configurations. See

Supplemental Material for a table of $R_C$ and $R_N$ for all possible permutations of Device N2 and Device N4.

In short, these transport signatures reflect the 1D nature of the sketched LaAlO$_3$/SrTiO$_3$ nanowires reported here. The superconducting critical temperatures for LaAlO$_3$/SrTiO$_3$ nanowires ($T_c = 100-300$ mK) [Table I] are comparable to those observed for bulk SrTiO$_3$ [15] and 2D LaAlO$_3$/SrTiO$_3$ heterostructures [10, 11]. However in the 1D limit, we observe a broader resistive transition, as well as non-zero resistance for $T \ll T_c$ [Fig. 2]. Broad superconducting transitions with decreasing temperature are observed in many other superconducting nanowires [1, 27, 28] and are typically attributed to thermally activated phase slips [29, 30] of the superconducting order parameter. Finite resistances well below the transition ($T \ll T_c$) are often considered to be signatures of quantum phase slips [1, 27, 28]. The non-linear behavior [Fig. 3] of the *V-I* curves for $I < I_c$ is similar to that of overdamped Josephson junctions where thermal noise [25, 31] and potentially quantum tunneling [25] cause phase slippage rates proportional to current, resulting in a nonlinear *V-I* in the superconducting state. Superconducting nanowires commonly exhibit these Josephson effects as well [25, 32].

The hysteresis in the *V-I* curve for Device N4 [Fig. 3] resembles an underdamped Josephson junction in a tilted washboard potential [26]. However, nanowires are typically overdamped because they have very small capacitance. The hysteresis could be caused by heating that occurs when the nanowire is driven normal [25, 32]. This interpretation may also help to explain the current drop above $I_c$ in Device N4.

Creation of nanowires with identical properties still presents technological challenges, similar to other superconducting nanowire materials. It is possible that differences arise from sensitivity to material defects or undetected fluctuations in writing parameters and AFM tip shape. Notably, these difference could result in variations in the carrier density along the nanowire that are detectable in transport experiments since both the spin-orbit strength and the superconducting order parameter vary with carrier density [11, 24]. Further work is under way to understand the role of magnetism and spin-orbit coupling in the superconducting state and how they may play a role in determining whether the pairing symmetry is *s*-wave or *p*-wave in 1D LaAlO$_3$/SrTiO$_3$ nanowires.

The effects of permuting current and voltages leads on the four-terminal d*V*/d*I* curves in Fig. 4 are significant. Current is sourced and voltages sensed across the same main channel, marked *L*, even though different leads are employed outside of this main channel. More surprising, it appears that key current-sourcing leads affect $R_C$ much more dramatically, for example Lead 2 in Device N2. In the normal state all three configurations are essentially identical. It appears that phase slips are much more readily produced by some current paths compared to others, while there is no intentional fabricated difference among the various leads.

The critical currents in these nanostructures are comparable in magnitude to the predicted step-like critical current in superconducting quantum point contacts $I_c = Ne\Delta_0 / \hbar$ where *N* is the number of channels [33]. This expression is applicable for $L < \xi_{SC}$. This is remarkable, given that in 2D [10], $\xi_{SC} \sim 100$ nm, which is two orders of magnitude smaller than *L* in these nanostructures. Another interesting observation is that $R_N$ and $R_C$ are generally of order ~$h/e^2$. A single-mode superconducting nanowire would essentially have the same properties [33, 34] as a superconducting quantum point contact.

The flexibility [20] of creating multiply-connected superconducting nanostructures make them attractive potential hosts for storing and manipulating quantum information [5, 8] or as sources of spin-entangled pairs of electrons [35]. For observation of Majorana states, several key ingredients are in place in this single material system for intrinsic topological superconductivity. Both the superconductivity [11] and spin-orbit coupling [24] in planar $LaAlO_3/SrTiO_3$ are tunable with electric field gating. Further study of $LaAlO_3/SrTiO_3$ nanostructures is needed to determine whether these more exotic states may be created, but $LaAlO_3/SrTiO_3$ nanostructures can readily be applied to study the fundamental physics of one-dimensional superconductivity in custom device geometries.

**References and Notes:**

† Present address: West Virginia University, Morgantown, WV 26506.
‡ Present address: Oak Ridge National Laboratory, Oak Ridge, TN 37831.
*Corresponding author: jlevy@pitt.edu

[1] K. Y. Arutyunov, D. S. Golubev, and A. D. Zaikin, Physics Reports **464**, 1 (2008).
[2] Y. Nakamura, Y. A. Pashkin, and J. S. Tsai, Nature **398**, 786 (1999).
[3] A. Y. Kitaev, Physics-Uspekhi **44**, 131 (2001).
[4] J. Alicea, Reports on Progress in Physics **75**, 076501 (2012).
[5] R. Lutchyn, J. Sau, and S. Das Sarma, Physical Review Letters **105**, 077001 (2010).
[6] J. D. Sau *et al.*, Physical Review Letters **104**, 040502 (2010).
[7] J. Alicea, Physical Review B **81**, 125318 (2010).
[8] V. Mourik *et al.*, Science **336**, 1003 (2012).
[9] L. Rokhinson, X. Liu, and J. Furdyna, Nature Physics - Advance online publication doi:10.1038/nphys2429 (2012).
[10] N. Reyren *et al.*, Science **317**, 1196 (2007).
[11] A. D. Caviglia *et al.*, Nature **456**, 624 (2008).
[12] A. Brinkman *et al.*, Nature Materials **6**, 493 (2007).
[13] Ariando *et al.*, Nature communications **2**, 188 (2011).
[14] L. Fidkowski *et al.*, http://arxiv.org/abs/1206.6959 (2012)
[15] J. Schooley, W. Hosler, and M. Cohen, Physical Review Letters **12**, 474 (1964).
[16] D. A. Dikin *et al.*, Physical Review Letters **107**, 056802 (2011).
[17] J. Bert *et al.*, Nature Physics **7**, 767 (2011).
[18] K. Michaeli, A. Potter, and P. Lee, Physical Review Letters **108**, 117003 (2012).
[19] C. Cen *et al.*, Nature materials **7**, 298 (2008).
[20] C. Cen *et al.*, Science **323**, 1026 (2009).
[21] K. Janicka, J. Velev, and E. Tsymbal, Physical Review Letters **102**, 106803 (2009).
[22] N. Reyren *et al.*, Applied Physics Letters **94**, 112506 (2009).
[23] P. Irvin *et al.*, Nature Photonics **4**, 849 (2010).
[24] A. D. Caviglia *et al.*, Physical Review Letters **104**, 126803 (2010).
[25] M. Tinkham *et al.*, Physical Review B **68**, 134515 (2003).

[26] M. Tinkham, *Introduction to Superconductivity* (McGraw-Hill, New York, 1995), 2 edn.
[27] N. Giordano, Physical Review Letters **61**, 2137 (1988).
[28] C. Lau *et al.*, Physical Review Letters **87**, 217003 (2001).
[29] J. Langer, and V. Ambegaokar, Physical Review **164**, 498 (1967).
[30] D. McCumber, and B. Halperin, Physical Review B **1**, 1054 (1970).
[31] V. Ambegaokar, and B. I. Halperin, Physical Review Letters **22**, 1364 (1969).
[32] P. Li *et al.*, Physical Review B **84**, 184508 (2011).
[33] C. W. Beenakker, and H. van Houten, Physical Review Letters **66**, 3056 (1991).
[34] C. Beenakker, Physical Review B **46**, 12841 (1992).
[35] G. B. Lesovik, T. Martin, and G. Blatter, European Physical Journal B **24**, 287 (2001).

**Acknowledgments**

We thank Lukasz Fidkowski, Roman Lutchyn, and Chetan Nayak for helpful discussions. This work is supported by AFOSR FA9550-10-1-0524 (J.L., C.B.E.), ARO W911NF-08-1-0317 (J.L.), NSF DMR-1104191 (J.L.), and DMR-0906443 (C.B.E).

| Device Name | $L$ (μm) | $R_N$ ($h/e^2$) | $R_C$ ($h/e^2$) | $T_c$ (mK) | $I_c$ (nA) | $I_c R_N$ (μV) |
|---|---|---|---|---|---|---|
| N1 | 6.5 | 4.6 | 0.01 | 150 | 1.5 | 178 |
| N2 | 10.0 | 0.4 | 0.07 | 110 | 7 | 71 |
| N3 | 10.0 | 1.1 | 0.25 | 180 | 2.8 | 75 |
| N4 | 6.5 | 0.8 | 0.05 | 220 | 5.6 | 116 |
| N5 | 6.5 | 1.2 | 0.02 | 135 | 2.5 | 75 |
| N6 | 2.0 | 0.4 | 0.06 | 125 | 3 | 37 |

TABLE I. Summary of nanostructured devices. The main channel length $L$ is the distance between the two transverse Hall crosses. All leads have width $w$~10 nm. $R_N$ and $R_C$ are resistances above and below $I_c$, respectively. $R_N$, $R_C$, and $I_c$ are representative values obtained from $V$-$I$ curves at T=50 mK.

**Figure Legends**

FIG. 1. (Color online) (a) Schematic for Device N2, which has five terminals, with $w$~10-nm leads and $L$=10-μm long main channel. (b) $V$-$I$ curves and (c) differential resistance curves showing the superconducting state. Complimentary curves in (b) and (c) show the superconducting state extinguished for sufficiently high temperature and magnetic field. The $V$-$I$ curve in (b) shows a finite slope with clear yet broadened switching current (~$I_c$) to the normal state (arrow). (c) The critical current $I_c$ is labeled at positive bias on the d$V$/d$I$ curve. The critical current is defined as the average current location of the two resistance peaks for positive ($I^+$) and negative ($I^-$) bias $I_c \equiv \frac{1}{2}\left(I^+ + |I^-|\right)$.

FIG. 2. (Color Online) d$V$/d$I$ plotted for several temperatures spanning the superconducting transition in Device N2. The transition is broad in temperature, with the gap opening incrementally as the temperature is lowered. Each curve is offset by log($T$), where $T$ is the temperature at which it was acquired.

FIG. 3. (Color Online) Normalized $V^{23}$-$I^{14}$ curves for three devices, below $T_c$. All three show that $V$-$I$ becomes non-linear as $I$ approaches $I_c$. Devices N2 and N3 have finite-width resistive transitions through $I_c$. Device N4 exhibits a characteristic 'bump' wherein the current evidently drops to a lower value during the superconducting to normal state transition. Inset: The bump is hysteretic in bias sweep direction, with a retrapping current $I_r$ occurring at lower voltage than $I_c$. ($V^{24}$-$I^{13}$). Device N2 and N3 have the same geometry, described in Fig. 1: five terminals; $L$=10 μm. Device N4 has six terminals; $L$=6.5 μm. The geometry is similar to that in Fig. 1(a), with the sixth lead wired to Lead 6 marked in Fig. 1(a) (not used in N2 and N3). ($T$=50 mK)

FIG. 4. (Color Online) Differential resistance for different permutations of current and voltage leads. d$V$/d$I$ is plotted for three unique permutations of Device N2. The graphical representations of the nanostructure illustrate the current pathways and voltage leads. Simply exchanging one current lead with one voltage lead changes the superconducting resistance, $R_C$, by more than a factor of 10 (red to blue). Switching the current path and fixing the voltage leads changes $R_C$ by a factor of 3 (red to green). ($T$=50 mK).

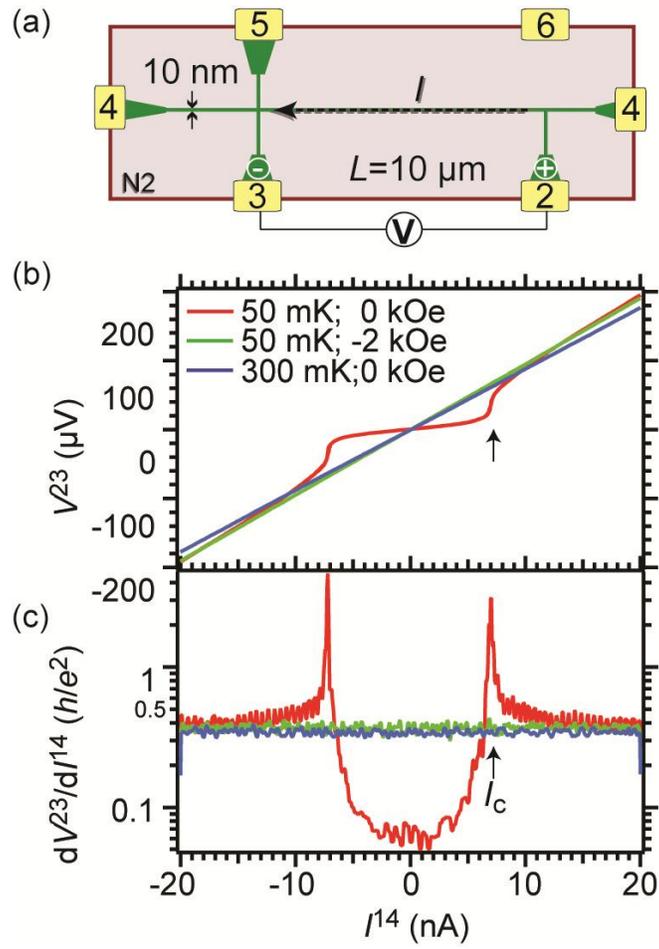

FIG. 1.

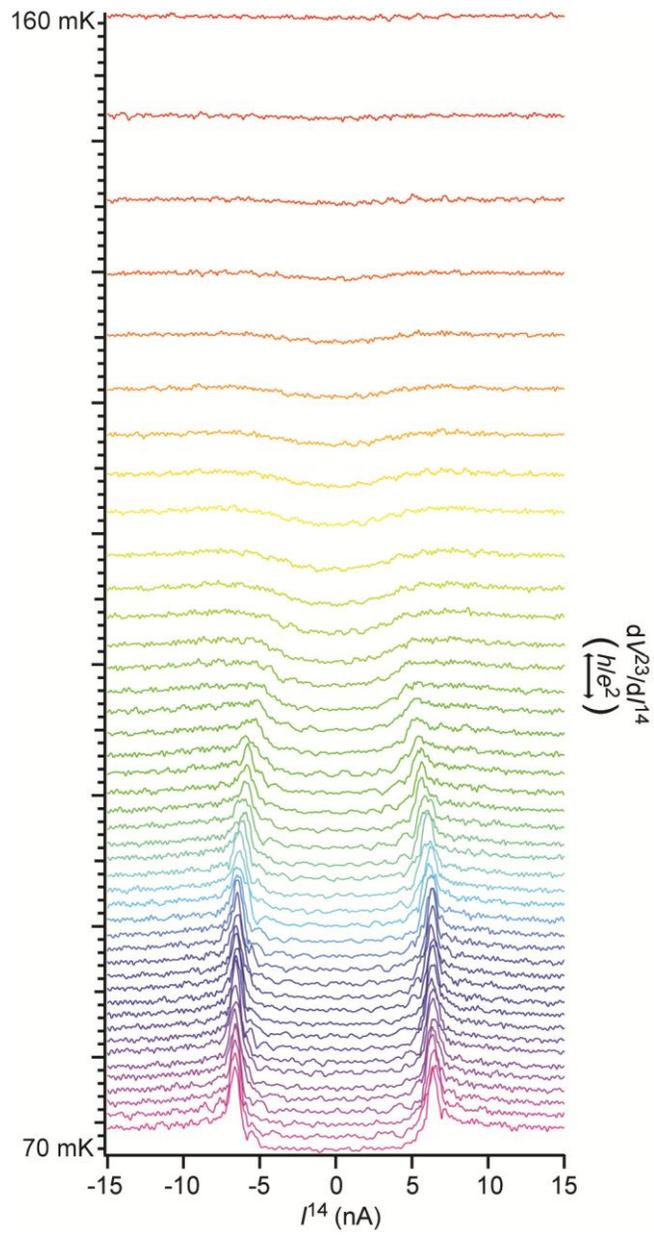

FIG. 2.

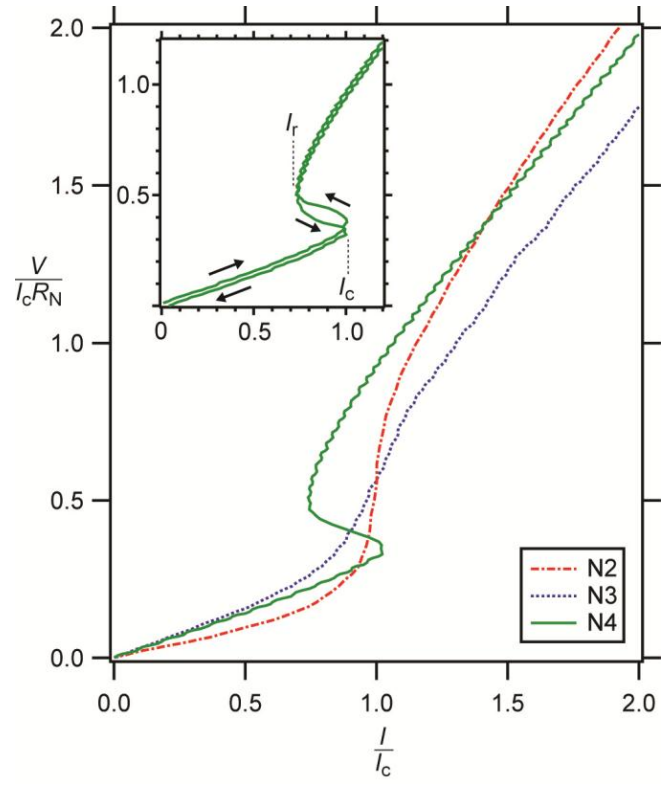

FIG. 3.

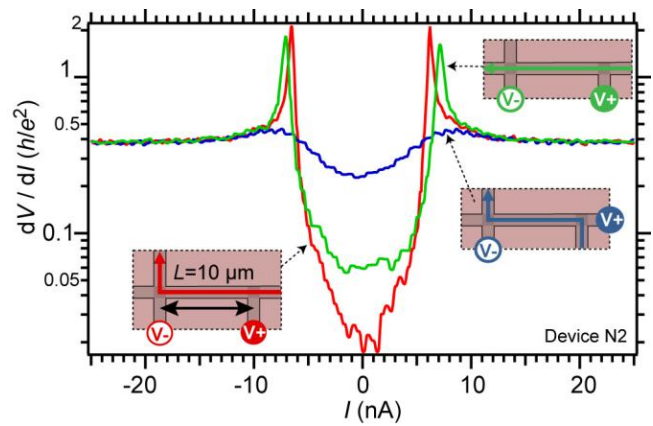

FIG. 4.

# Supplemental Material
# Superconducting LaAlO$_3$/SrTiO$_3$ Nanowires

Joshua P. Veazey[1], Guanglei Cheng[1], Patrick Irvin[1], Cheng Cen[1,†], Daniela F. Bogorin[1,‡], Feng Bi[1], Mengchen Huang[1], Chung-Wung Bark[2], Sangwoo Ryu[2], Kwang-Hwan Cho[2], Chang-Beom Eom[2] and Jeremy Levy[1,*]

[1]*Department of Physics and Astronomy, University of Pittsburgh, Pittsburgh, Pennsylvania 15260, USA,* [2]*Department of Materials Science and Engineering, University of Wisconsin-Madison, Madison, Wisconsin 53706, USA.*

† Present address: West Virginia University, Morgantown, WV 26506.
‡ Present address: Oak Ridge National Laboratory, Oak Ridge, TN 37831.
*Corresponding author: jlevy@pitt.edu

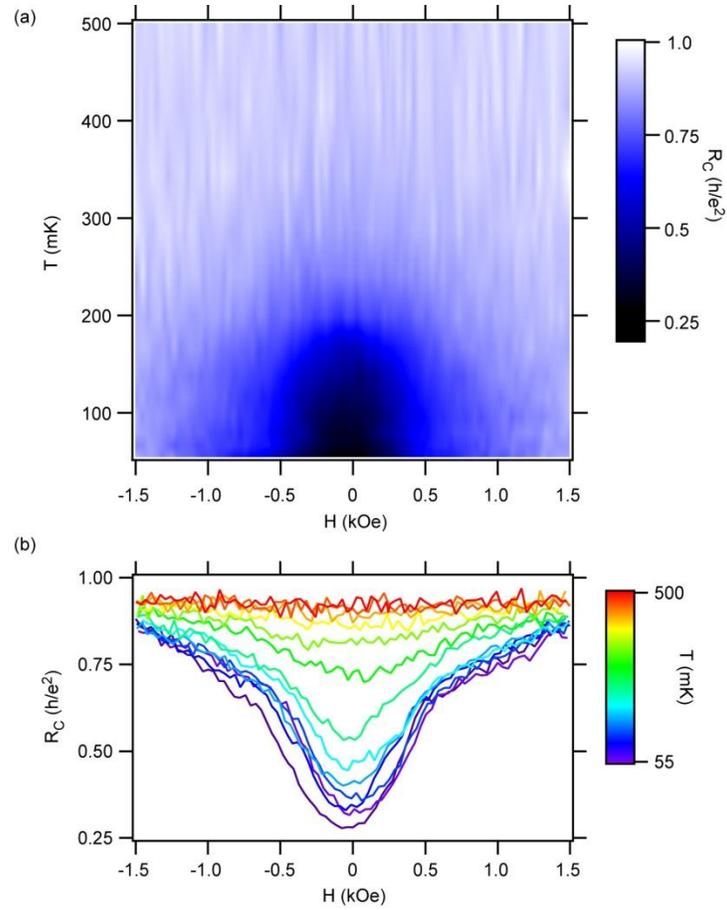

FIG. SM1. (a) Temperature and magnetic field phase diagram of Device N3. The resistance reaches a minimum, in the superconducting phase. The normal state ($R_N \sim h/e^2$) is restored for temperatures above T~200 mK and fields H~1.5 kOe. (b) vs. magnetic field, $H$, for several temperatures above and below $T_c$.

# Permutations of the current- and voltage-sensing leads

Longitudinal ($R_{xx}$) resistances measured in the normal and superconducting states depend upon the current and voltage leads used to couple to the main channel of the Hall bar. The resistances of both states are listed for each of the longitudinal lead-permutations of Device N2 (five leads) and Device N4 (six leads) in Tables SMI and SMII, respectively. The ratio of the two resistances is also given for each permutation. For Device N2, $R_C$ varies by an order of magnitude, while $R_N$ remains relatively constant. For Device N4, $R_C$ again varies by more than an order of magnitude, in some cases approaching zero. $R_N$ for Device N4 shows more moderate (30%) variation.

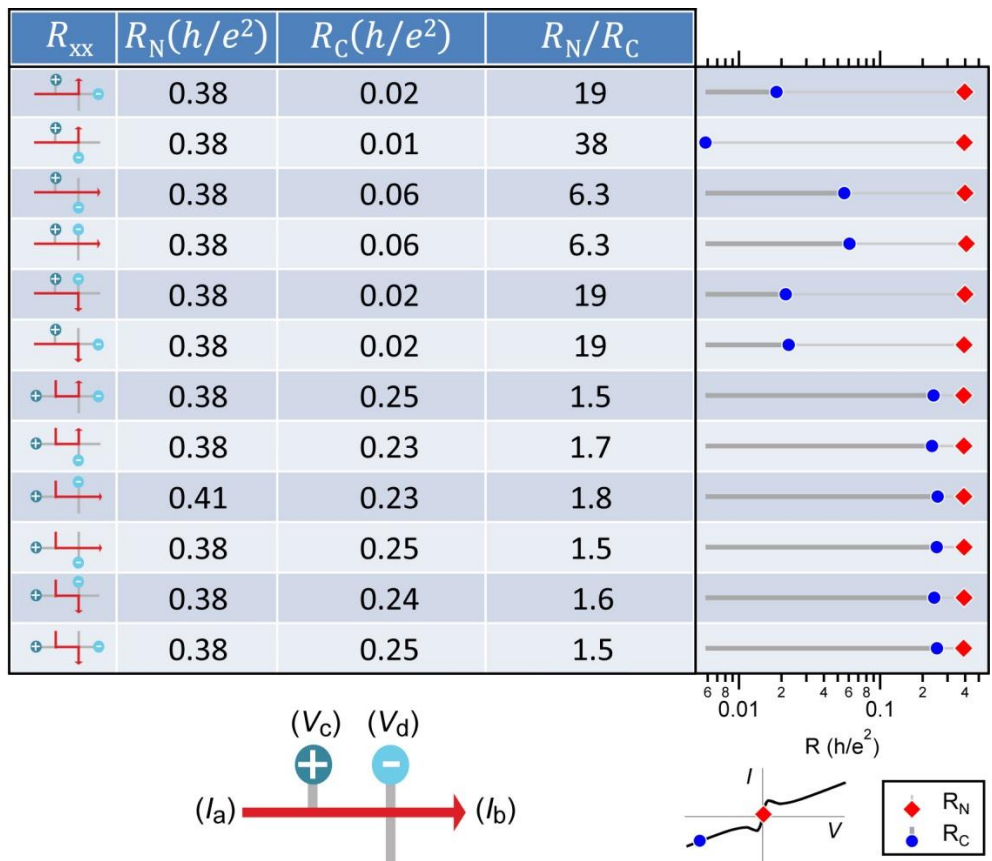

| $R_{xx}$ | $R_N (h/e^2)$ | $R_C (h/e^2)$ | $R_N/R_C$ |
|---|---|---|---|
|  | 0.38 | 0.02 | 19 |
|  | 0.38 | 0.01 | 38 |
|  | 0.38 | 0.06 | 6.3 |
|  | 0.38 | 0.06 | 6.3 |
|  | 0.38 | 0.02 | 19 |
|  | 0.38 | 0.02 | 19 |
|  | 0.38 | 0.25 | 1.5 |
|  | 0.38 | 0.23 | 1.7 |
|  | 0.41 | 0.23 | 1.8 |
|  | 0.38 | 0.25 | 1.5 |
|  | 0.38 | 0.24 | 1.6 |
|  | 0.38 | 0.25 | 1.5 |

TABLE SMI. Longitudinal resistances for multiple permutations of Device N2. $R_C$ is the resistance in the superconducting phase, $R_N$ is the resistance in the normal phase, and the ratio is $R_N/R_C$. Different lead-permutations produce changes in the normal and superconducting resistances. The key below the table indicates current and voltage leads in the compact permutation icon used in the table.

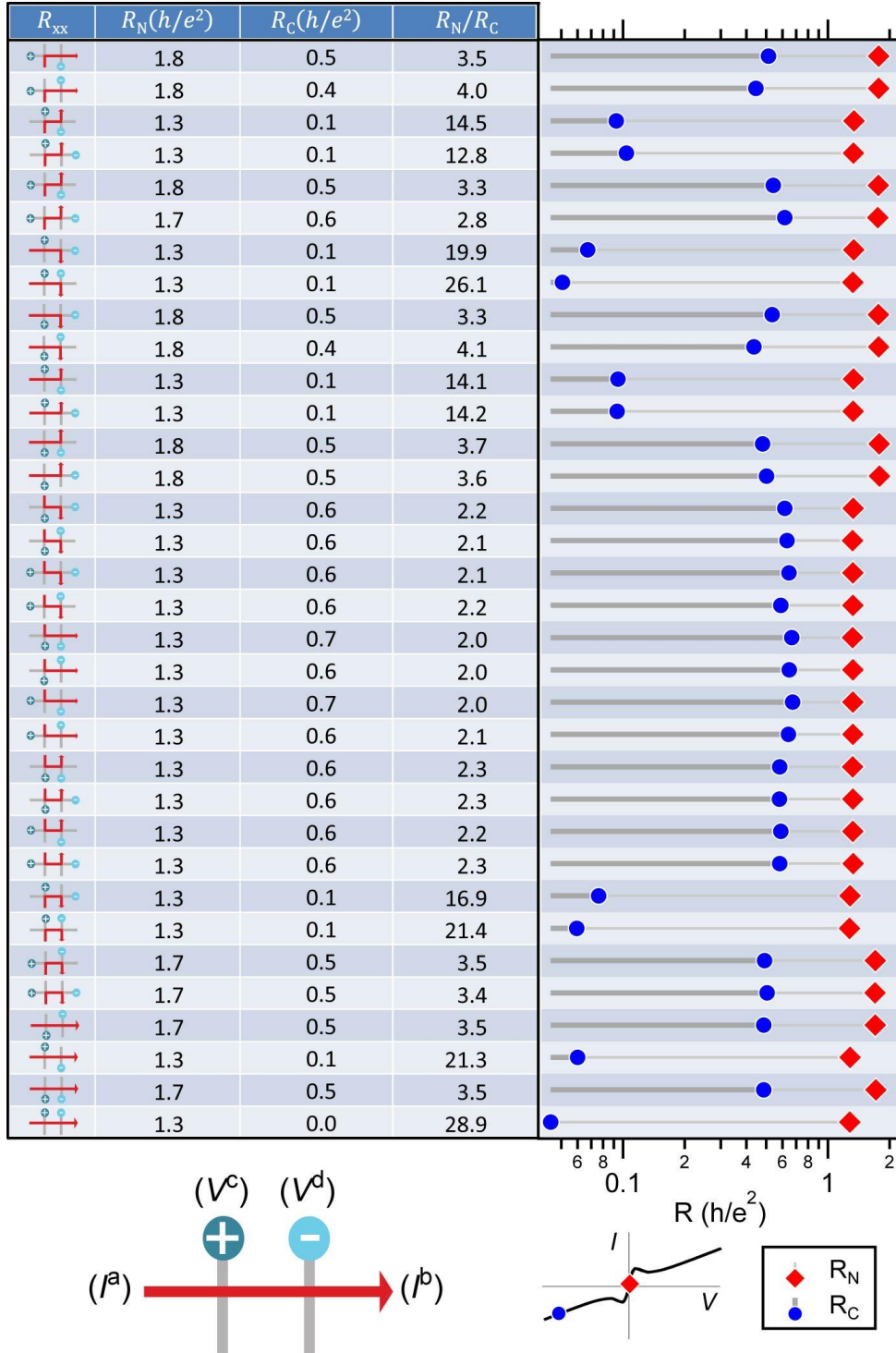

TABLE SMII. Longitudinal resistances for multiple permutations of Device N4. $R_C$ is the resistance in the superconducting phase, $R_N$ is the resistance in the normal phase, and the ratio is $R_N/R_C$. Different lead-permutations produce changes in the normal and superconducting resistances. The key below the table indicates current and voltage leads in the compact permutation icon used in the table.